# ENHANCED CLASSIFICATION OF ORAL CANCER USING DEEP LEARNING TECHNIQUES

Dr. Senthil Pandi S
Associate Professor
Rajalakshmi Engineering College
Chennai, India
senthilpandi.s@rajalakshmi.edu.in

Hirthik Mathesh GV
Computer Science and Engineering
Rajalakshmi Engineering College
Chennai, India
hirthikmatheshgv@gmail.com

Kavin Chakravarthy M
Computer Science and Engineering
Rajalakshmi Engineering College
Chennai, India
kavinchakravarthy1021@gmail.com

*Abstract— Oral cancer constitutes a significant global health concern, resulting in 277,484 fatalities in 2023, with the highest prevalence observed in low- and middle-income nations. Facilitating automation in the detection of possibly malignant and malignant lesions in the oral cavity could result in cost-effective and early disease diagnosis. Establishing an extensive repository of meticulously annotated oral lesions is essential. In this research photos are being collected from global clinical experts, who have been equipped with an annotation tool to generate comprehensive labelling. This research presents a novel approach for integrating bounding box annotations from various doctors. Additionally, Deep Belief Network combined with CAPSNET is employed to develop automated systems that extracted intricate patterns to address this challenging problem. This study evaluated two deep learning-based computer vision methodologies for the automated detection and classification of oral lesions to facilitate the early detection of oral cancer: image classification utilizing CAPSNET. Image classification attained an F1 score of 94.23% for detecting photos with lesions 93.46% for identifying images necessitating referral. Object detection attained an F1 score of 89.34% for identifying lesions for referral. Subsequent performances are documented about classification based on the sort of referral decision. Our preliminary findings indicate that deep learning possesses the capability to address this complex problem.*

## I. INTRODUCTION

In recent years oral cancer became a major issue in healthcare, oral cancer is distinguished by its high death and morbidity rates as well as its delayed detection. In 2021, GLOBOCAN projected 234,384 fatalities and 387,864 new cases. Three-quarters of the worldwide Oral cancer is more common in low-income people. Various nations, with Africa and India accounting for half of all cases. The two main risk factors for the disease are excessive alcohol use and tobacco use in any form. Chewing betel quid, which is often made up of limes, nut and leaf and maybe tobacco, is a factor that is most prevalent in South and North. These days, thanks to aggressive marketing tactics, these quads are widely accepted by the public and sold commercially in sachets. In LMICs, where over two-thirds of cases appear at advanced stages and have low survival rates, oral cancer is usually linked to late presentation. Cancer treatment is exceedingly expensive, particularly when it is late in the disease's progression. One major factor contributing to late identification of oral cancer is the general lack of awareness and the ignorance of medical professionals. This type of cancer is frequently preceded by visible oral lesions known as Threatening malignant disorders, which can be found during general checkup by a clinical oral examination (COE) conducted by a general dentist. Therefore, a late diagnosis need not be a defining characteristic. The patient is directed to a specialist for diagnosis confirmation and additional treatment if a worrisome lesion is found. According to earlier research conducted in India, screening has reduced mortality among alcohol and tobacco users, led to early diagnosis, and down staged the disease. Due to a lack of specialists and health resources, LMICs bear the majority of the burden of oral cancer, hence screening programs must provide an affordable and effective method of detection. Using telemedicine would be one such practical strategy. When comparing the clinical diagnoses provided by professionals conducting a COE to those made after seeing mobile phone photos, demonstrated a moderate to high concordance. Specialists' virtual consultations could increase screening programs' referral accuracy. It would be very advantageous to advance this idea by adding a detection system connected to deep learning method to examine photos from mobile phones.

The research community is increasingly using various advanced algorithms, to enhance the functionality of computer-aided medical diagnostic systems. Histopathological images, confocal laser endomicroscopy (CLE) images, hyperspectral images, autofluorescence images (AFI), and fluorescence images—which usually show close-ups of the oral lesions—are among the various image types used in recent studies to identify oral cancer using deep learning algorithms. A method that combines the output of three pre-trained DL models using a transfer learning technique to categorize histopathological pictures into two groups: normal and malignant. In order to categorize CLE pictures as either normal or malignant, In one of the methods evaluated a deep Convolutional Neural Network (CNN) based on LeNet5, and the findings demonstrated that it performed better than feature-based classification techniques. In [14], a regression-based partitioned deep CNN was used to classify multidimensional hyperspectral pictures into three categories: normal, benign, and malignant. Additionally, a novel methodology created a mobile phone-connected device that can capture fluorescent oral images. They then classified the photos into two classes using state-of-the-art (SOTA) deep neural networks. All of these methods, however, need expensive equipment or a particularly designed screening platform to capture the ROI at a high magnification, which prevents many people from using them. While a new research employed higher order techniques, Imaging, and laws texture energy, one of the authors used the grey scale methodology, and the first papers in the field concentrated on texture-based characteristics. Deep learning, which is the use of artificial neural networks with multiple layers of neurons that rely on massive datasets and quick processing capacity to understand intricate patterns, has become increasingly popular in the more recent studies. More precisely, the deep convolutional neural network (CNN) used in these publications explicitly assumed that the inputs were images in their architectures. CNNs have become increasingly popular in the field of computer vision after AlexNet won the ImageNet picture classification competition in 2017.

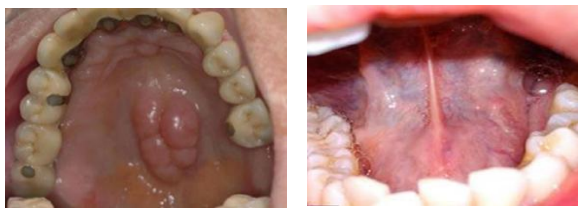

**Fig.1. Various Oral cavity Images from different Datasets.**

## II. LITERATURE SURVEY

In [1] Roshan et.al, proposed a method for automating the detection of tumors that may be cancerous could result in an early and inexpensive diagnosis. This research proposes a novel approach for automated systems that combines deep neural networks with bounding box annotations from several doctors. Two computer vision techniques based on deep learning were evaluated for automated oral lesion detection and categorization. An F1 score of 87.07% was obtained for lesions identification in image classification, and 78.30% was obtained for referral identification.

Oral carcinoma cells in represents a significant health issue [2], and prompt identification is essential for effective intervention and enhanced survival rates. Prior research has predominantly depended on pictures for the classifying various types of lesions in oral cavity, neglecting the advantages of integrating several modalities. This paper presents a multimodal deep-learning framework that utilizes many data sources, including patient metadata, to replicate physicians' diagnostic methods for the early diagnosis of oral cancer. The proposed method attained an overall accuracy of 87%, precision of 83%, recall of 83%, F1-score of 88%, and a Matthews Correlation Coefficient of 0.75.

In [3] A novel approach has been introduced to distinguish between carcinoma and non-carcinoma cells and to categorize their initial phases utilizing a Deep learning technique known as Ada Boosting. The approach employs five unique colour spaces to extract colour and texture information, subsequently classifying them with a Light GBM classifier. The overall performance is encouraging, exhibiting a testing accuracy of 97.25%, precision of 98.23%, recall of 98.41%, f1-score of 96.24%, and specificity of 98.41% for binary classification, alongside 99.21% accuracy, 99.76% precision, 98.72% recall, 99.48% f1-score, and 98.043% specificity for multi-class classification.

In [4] Oral cancer constitutes a worldwide health hazard, and prompt diagnosis is essential for effective improving the rate of survival. Histopathological process is manually done and is labour-intensive and susceptible to subjective variations. This study investigates the capabilities of artificial intelligence (AI) in diagnosing OSCC through three approaches: Reset, Cat Boost and Gabor method. The third technique, integrating Gabor filtering, exhibited remarkable performance, attaining an accuracy of 94.92%, precision of 95.51%, sensitivity of 84.30%, specificity of 95.54%, F1 score of 94.90%, and AUC of 94.9%.

In [5] This study employs deep learning methodologies to examine histological images of oral cells for the identification of oral carcinoma The suggested paradigm employs transfer learning and ensemble learning in two distinct phases. Initially, multiple Deep Neural Network (CNN) models are evaluated for OSCC identification. In the subsequent phase, an ensemble model is developed utilizing the two most effective pre-trained CNNs. The suggested classifier is evaluated against prominent models and exhibits its efficacy. Following a three-phase comparative examination, the ensemble classifier demonstrates improved performance with an accuracy of 97.88%.

In [6] The human brain and oral cavity are intricately connected, with the mouth and face housing 45-60% of the body's sensory and motor nerves. Recognizing orofacial manifestations of neurological illnesses is vital for dental surgeons, and comprehending these manifestations is imperative for prompt diagnosis and intervention. The new techniques technology facilitates the secure process of medical data, essential for disease diagnosis and treatment. This paper presents a novel methodology, which gathers dental data from various hospitals pertaining to the oral cavity and central nervous system, extracts features, and employs the IGPLONN algorithm to enhance prediction accuracy. The hybrid optimized technique enhances detection rates for oral-related neurological diseases and regulates predictive indicators for dental metastases. The experimental assessment conducted on MATLAB validated the system's superiority, achieving a maximum accuracy of 98.3%.

This study evaluates the impact of Deep learning [7] on enhancing oral cancer detection. Oral cancer is a significant global health issue, frequently diagnosed at advanced stages, resulting in unfavourable prognoses. Artificial intelligence methodologies, especially Novel algorithms, can precisely evaluate digital images and histopathological slides, assisting clinicians in risk evaluation and early detection. IoT-enabled devices facilitate real-time monitoring and surveillance, permitting the initial detection of oral cancer indicators. Image processing methods and big data analytics improve early detection. Validation, data protection, and regulations are essential for effective clinical application. Future efforts involve developing multimodal imaging techniques and incorporating them into population-based screening programs.

Identifying oral squamous cell cancer is a considerable problem because to delayed diagnosis and costly data procurement. An economical, automated screening system is [8] essential for early illness identification and integration with vital sector applications. Explainable Artificial Intelligence (XAI) offers assistance; nonetheless, existing methodologies are predominantly data-driven. A solution integrating Case-Based Reasoning with Informed Deep Learning (IDL) is provided. The IDL methodology addresses data defects, such as labelling inaccuracies and artifacts, attaining an accuracy of 85%, which exceeds the 77% attained by DL independently. The human-centred explainability of both approaches is also assessed.

This study investigates the application of deep learning (DL)[9] for diagnosing and forecasting oral cancer prognosis through Deep neural networks (DNNs) and Long Short Term Memory (LSTM) over the last five years. It underscores the significance of various oral cancer datasets. Nonetheless, deep learning methods encounter limitations such as input variability and model interpretability, necessitating additional focus to fully exploit deep learning's potential in oral cancer treatment.

This study sought to forecast the short-term efficiency and recurrence of oral cancer [10] following aminolaevulinic acid photodynamic treatment (ALA-PDT). The study employed regression models to develop predictive models for total rate of response, full response rate. An autoencoder utilizing deep learning was employed to extract features from pathology sections and integrate clinical factors to enhance predictive performance. The research identified a correlation between enlarged cells in the recurrence following PDT. The integration of clinical factors and machine learning enhanced the efficacy of the model by more than 40%.

The research assessed the application of computer vision models [11] for the classification and identification of normal oral mucosa, and various prevalent oral disorders in clinical oral images. The research utilized 625 clinical oral pictures, categorized into 251 different aphthous ulcers (RAU). Four CNN models were employed for detection tasks, while three were utilized for classifying the diseases tasks. The Pretrained ResNet101 showed the good performance in classification with a precision of 90.16%, recall of 93.14%, F1 score of 93.44%.

Oral cancer is a common affliction of the oral cavity, and comprehending its initial phases is essential for healthcare professionals [12]. Computer-aided diagnostic methods utilizing deep learning have demonstrated efficacy in identifying oral cancer lesions. Transfer learning can effectively manage extensive training datasets in biomedical image classification. This study presents two methodologies for the classification of carcinoma cell histopathological pictures. The initial method employs transfer learning-enhanced deep convolutional neural networks (DCNNs) to distinguish between benign and malignant tumors. The second approach suggests a foundational DCNN architecture of 10 convolutional layers. Experimental findings indicate that ResNet50 surpasses fine-tuned models and the CNN model, achieving an accuracy of 95.4%, precision of 98%, and recall of 93%.

The research sought to assess the likelihood of carcinoma in oral cancer lesions with Deep Learning techniques [13]. Two hundred sixty-one lesions were examined utilizing normal digital photos. A deep algorithm pipeline, comprising U-Net segmentation and a multi-task RNN classifier, was employed to identify the risk of carcinoma. A heatmap for explainability was generated with LIME to elucidate the model's decisions. The model attained a Dice coefficient of 0.561, a sensitivity of 1, a specificity of 0.692, and a specificity of 0.760 with a sensitivity of 0.938. The model's output exhibited significant confidence and clarity of interpretation.

Anatomical pathology is experiencing a third revolution, shifting from antilog to digital pathology [14] while integrating artificial intelligence technology. Predictive models are becoming influential as they can enhance diagnostic procedures and laboratory operations, reducing consumable consumption and turnaround time. A novel algorithm was designed to produce synthetic Ki-67 immunohistochemistry from Haematoxylin-stained pictures. The model was developed utilizing 175 carcinoma cells specimens from the archives of the Pathology Unit at University Federico II. The model generated lifelike synthetic visuals and attained elevated positive prediction values. This model is an effective instrument for obtaining Ki-64 data directly from differential slides, minimizing laboratory requirements and enhancing patient care.

The prognosis for oral cancer is unfavourable, with more than fifty percent identified at advanced stages [15]. Historically, screening approaches depended on the clinical experience of health workers, and there is no standardized approach for detecting individuals with oral cancer cells through camera images. A deep learning system was created utilizing deep neural networks to identify cancer from photographic pictures. The algorithm was trained on a randomly chosen segment of the development dataset and evaluated on an internal validation dataset. The algorithm's efficacy was evaluated against various coral cavity cancer dataset. The method attained an accuracy of 0.934 on the internal validation dataset.

III. PROPOSED METHODOLOGY

One of the most prevalent malignancies in the world, oral cancer is distinguished by its high death and morbidity rates as well as its delayed detection. In 2021, GLOBOCAN projected 234,384 fatalities and 387,864 new cases. Three-quarters of the worldwide Oral cancer is more common in low-income people. Various nations, with Africa and India accounting for half of all cases. The two main risk factors for the disease are excessive alcohol use and tobacco use in any form. Chewing betel quid, which is often made up of limes, nut and leaf and maybe tobacco, is a factor that is most prevalent in South and North. These days, thanks to aggressive marketing tactics, these quads are widely accepted by the public and sold commercially in sachets. In LMICs, where over two-thirds of cases appear at advanced stages and have low survival rates, oral cancer is usually linked to late presentation. Cancer treatment is exceedingly expensive, particularly when it is late in the disease' progression. One major factor contributing to late identification of oral cancer is the general lack of awareness and the ignorance of medical professionals. This type of cancer is frequently preceded by visible oral lesions known as Threatening malignant disorders, which can be found during general checkup by a clinical oral examination (COE) conducted by a general dentist. Therefore, a late diagnosis need not be a defining characteristic. The patient is directed to a specialist for diagnosis confirmation and additional treatment if a worrisome lesion is found. According to earlier research conducted in India, screening has reduced mortality among alcohol and tobacco users, led to early diagnosis, and down staged the disease. Due to a lack of specialists and health resources, LMICs bear the majority of the burden of oral cancer, hence screening programs must provide an affordable and effective method of detection. Using telemedicine would be one such practical strategy. When comparing the clinical diagnoses provided by professionals conducting a COE to those made after seeing mobile phone photos, demonstrated a moderate to high concordance. Specialists' virtual consultations could increase screening programs' referral accuracy. It would be very advantageous to advance this idea by adding a detection system connected to deep learning method to examine photos from mobile phones. The research community is increasingly using Deep learning and machine learning (ML) algorithms, to enhance the functionality of computer-aided medical diagnostic systems. Histopathological images, confocal laser endomicroscopy (CLE) images, hyperspectral images, autofluorescence images (AFI), and fluorescence images—which usually show close-ups of the oral

lesions—are among the various image types used in recent studies to identify oral cancer using deep learning algorithms. A method that combines the output of three pre-trained DL models using a transfer learning technique to categorize histopathological pictures into two groups: normal and malignant. In order to categorize CLE pictures as either normal or malignant, in one of the methods evaluated a deep Convolutional Neural Network (CNN) based on LeNet5, and the findings demonstrated that it performed better than feature-based classification techniques. In [14], a regression-based partitioned deep CNN was used to classify multidimensional hyperspectral pictures into three categories: normal, benign, and malignant. Additionally, a novel methodology created a mobile phone-connected device that can capture fluorescent oral images. They then classified the photos into two classes using state-of-the-art (SOTA) deep neural networks. All of these methods, however, need expensive equipment or a particularly designed screening platform to capture the ROI at a high magnification, which prevents many people from using them. While new research employed higher order techniques, Imaging, and laws texture energy, one of the authors used the grey scale methodology, and the first papers in the field concentrated on texture-based characteristics. Deep learning, which is the use of artificial neural networks with multiple layers of neurons that rely on massive datasets and quick processing capacity to understand intricate patterns, has become increasingly popular in the more recent studies. More precisely, the deep convolutional neural network (CNN) used in these publications explicitly assumed.

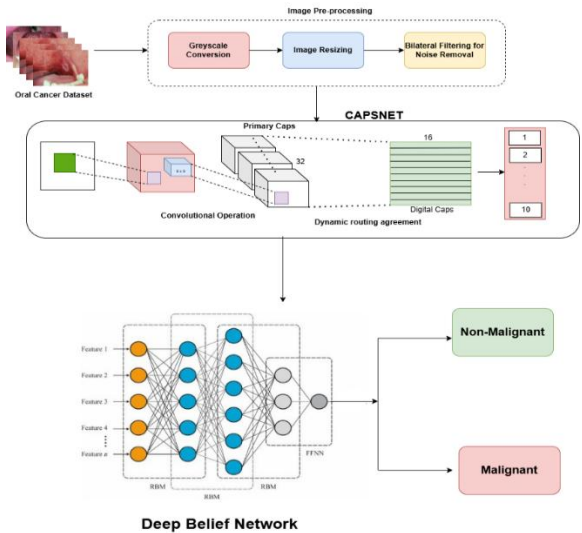

**Fig 2. Hybrid Model using CAPSNET and DBT for Oral Cancer Detection.**

## IV. IMAGE PRE PROCESSING

The image patches obtained from two distinct databases exhibit differences in colour and shade, likely attributable to heterogeneity in circumstances and staining processes. To mitigate the impact of disparate colour staining and render the model independent of colour attributes, error standardization has been implemented for the training set. Subsequently, a median filter was employed for additional reducing noise, as specific light and dark pixels are linked to the microscope during picture acquisition. The gathered picture patches of benign and malignant instances are inadequate, hence insufficient for generalizing the Deep learning models for classification. Thus, data augmentation is employed to enhance the quantity of picture patches, thereby mitigating the overfitting problem and ensuring better generalization of the Deep learning models during evaluation. Data augmentation enables the generation of a substantially greater quantity of data for training each class, involving image alteration using various ways.

## V. CAPSNET

The human brain contains specialized processing units known as "capsules," which are highly effective at interpreting various visual inputs and encoding, deformation, velocity, albedo, colour, and texture. For efficient processing, the brain must be able to route low-level visual information to the most relevant capsule. In essence, a capsule is a structured set of neuronal layers, where additional layers are not simply stacked but rather embedded within existing layers. In a conventional neural network, layers are continuously added in sequence, while in CapsNet, layers are organized hierarchically within a single "capsule." Each capsule represents a unique entity within an image, and the state of neurons in the capsule encodes that entity's specific characteristics. A capsule outputs a vector that indicates the presence of the entity, where the direction of the vector reflects the entity's properties. This vector is sent to all potential "parent" capsules within the network. For each parent capsule, a prediction vector is calculated by multiplying the output vector by a weight matrix. The parent capsule with the highest agreement—measured by the scalar product of its own vector and the prediction vector—strengthens its connection with the original capsule, while other connections weaken. This process, known as "routing by agreement," is more effective than traditional methods like max-pooling, which routes based on the most prominent feature found in the lowest layer. In addition to dynamic routing, CapsNet introduces a "squashing" function within each capsule. Squashing is a type of non-linearity applied to the vector output, ensuring that each vector is scaled to reflect the entity's presence without magnifying it excessively. Unlike CNNs, which apply activation functions to each individual layer, CapsNet applies the squashing function to the combined vector output from each capsule, enabling more accurate and efficient representation of visual entities.

## VI. DEEP BELIEF NETWORK

Initially all the new data input are whitened to reduce relationship between adjacent threshold levels. This process, called "Improvement of images," involves displaying the input data onto matrices and standardizing them so that all threshold values achieve a variance of 0. The resulting enhanced image is then directed to the visible layer of the Deep Network and is randomly divided into overlapping small groups. Mini-batching further dividing the input data into manageable segments and thereby applying the proposed algorithm, helping to reduce noise in cases where data samples lack representativeness. The Deep Belief Network (DBN) uses a hierarchical framework to separate basic features from advanced ones. A Deep Boltzmann Machine (DBM) is a generative model that stacks Convolutional Restricted Boltzmann Machines (CRBMs) vertically. Each unit contains a hidden layer (H), a input layer (I), and a pooling layer. The visible layer is represented by an RK * RK matrix of binary units, forming the image's visible layer. The total number of hidden units is k times $N^2$ where (H) contains(N) groups each comprised of MQ * MQ binary units. In the input layer, each unique group is assigned a window of size MN * MN (where (MN = MR - MQ + 1). A weight matrix, denoted as W, establishes symmetric relationships between hidden and visible units, which can also be thought of as filters. Additionally, there are N elements in the pooling layer, each containing NP * NP units. An equal factor C is used by various layers to reduce the dimensionality of (H)'s representation. To achieve this, the max layer selects the highest values from (Q * Q) windows in the hidden layer. This max-pooling operation improves the upper layers' stability against small input changes while lowering overall computational cost. For feature extraction from neuroblastoma histology images, a three-layer DBN is used. This DBN has several key hyperparameters that need tuning. The first parameter is the total number of neurons. While fewer hidden layers reduce training time, the network's performance may decline due to an inability to ignore basic noise in the input images. Increasing various layers in the model can improve performance but risks overfitting, higher computational demands, and longer training times. The new element is the number of groups within hidden layers, which lacks a universal selection rule. Adjusting this parameter affects both learning time and performance, similar to changes in the number of hidden layers. The third hyperparameter is the mini-batch size, which can reduce the computational load of the DBN network.

## VII. RESULTS AND DISCUSSION

This section examines the results of the developed method, including both for testing and training the model. The AUC curve is obtained from the training set, reflecting the model's learning progress, while the validation curve, based on a validation set, illustrates the model's generalization ability. The training loss represents the result on the input dataset, and the testing loss is the error after applying the executed network to the validation dataset. Initially, the study conducted trials for 10 epochs, which were later extended to 30 epochs. This choice of 30 epochs was made based on extensive findings showing that learning generally converges within this period across tests, with the exception of the baseline DNN model. Figure 2 presents the training accuracy and loss curves for the proposed method. The data reveals an exponential rise in model accuracy up to the 12th epoch, after which it begins to plateau. Both the training and validation loss curves decrease, indicating a trend toward stability. The diminishing loss signifies that the model is effectively capturing relevant features, though accuracy plateaus around 93.5% and 94.55%, respectively. Accuracy increases up to the 8th epoch, stabilizing as they reach a plateau. Training is halted at 12 epochs after achieving peak validation accuracy of 97.1%. The validation loss curve drops sharply in the initial epoch, then stabilizes at 0.1520 after the 8th epoch. The model continues to increase in training and validation accuracy up to 50 epochs, stabilizing at 94.28% and 94.55%, respectively, indicating strong model fit to the training data. The validation loss curve closely follows the trend of the training loss, with final loss values of 0.16543 for validation and 0.18432 for training, confirming the model's stability and alignment across both datasets.

**Table 1. Various Image classes and their Precision, Recall and F1 score for the given dataset**

| Image Class | Precision (%) | Recall (%) | F1 score (%) |
|---|---|---|---|
| Lesion not found | 90.86 | 91.23 | 80.65 |
| Image with no referral | 93.26 | 90.21 | 94.52 |
| Visited for different Reasons | 89.32 | 91.24 | 80.15 |
| Low Risk of Cancer | 90.88 | 89.23 | 87.21 |
| High Risk of Cancer | 94.24 | 90.21 | 84.21 |

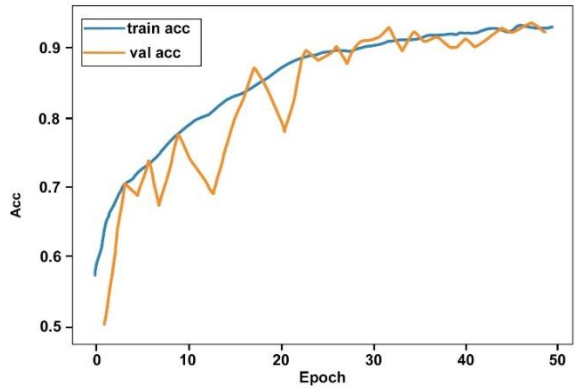

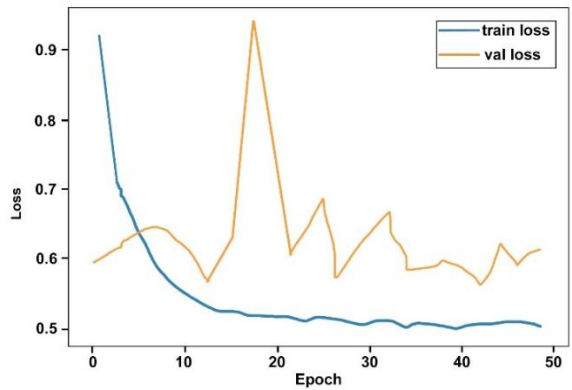

**Fig 3. Train and Validation Accuracy for the Proposed model**